\documentstyle[prd,aps,floats,twocolumn]{revtex}

\def\S{{\cal S}}
\def\be{\begin{equation}}
\def\ee{\end{equation}}
\def\bea{\begin{eqnarray}}
\def\eea{\end{eqnarray}}

\renewcommand\({\left(}
\renewcommand\){\right)}
\renewcommand\[{\left[}
\renewcommand\]{\right]}

\newcommand\eq[1]{Eq.~(\ref{#1})}
\newcommand\eqs[2]{Eqs.~(\ref{#1}) and (\ref{#2})}

\newcommand\eqst[2]{Eqs.~(\ref{#1})--(\ref{#2})}

\newcommand\GeV{\,\mbox{GeV}}
\newcommand\MeV{\,\mbox{MeV}}
\newcommand\keV{\,\mbox{keV}}

\newcommand\lsim{\mathrel{\rlap{\lower4pt\hbox{\hskip1pt$\sim$}}
    \raise1pt\hbox{$<$}}}
\newcommand\gsim{\mathrel{\rlap{\lower4pt\hbox{\hskip1pt$\sim$}}
    \raise1pt\hbox{$>$}}}

\newcommand\diff{\mbox d}

\def\calp{{\cal P}}
\def\calr{{\cal R}}

\newcommand\bfx{{\bf x}}

\newcommand\sub[1]{_{\rm #1}}

\begin{document}
\preprint{PU-ICG-02/15, astro-ph/0208055v2} 
\draft

%
%
\input epsf
\renewcommand{\topfraction}{0.99}
\renewcommand{\bottomfraction}{0.99}
\twocolumn[\hsize\textwidth\columnwidth\hsize\csname@twocolumnfalse\endcsname

\title{The primordial density perturbation in the curvaton scenario}
\author{David H.\ Lyth$^1$, Carlo Ungarelli$^2$ and David Wands$^2$}
\address{(1) Department of Physics, Lancaster University, Lancaster
LA1 4YB,~~~U.~K.}
\address{(2) Institute of Cosmology and Gravitation, University of
Portsmouth, Portsmouth PO1 2EG,~~~U.~K.}
\date{\today}
\maketitle
\begin{abstract}
  We analyse the primordial density perturbation when it is generated
  by a `curvaton' field different from the inflaton.  In some cases this
  perturbation may have large isocurvature components, fully
  correlated or anti-correlated with the adiabatic component.  It may
  also have a significant non-Gaussian component.
  All of these effects are calculated in a form which will enable direct
  comparison with current and forthcoming observational data.
\end{abstract}

\pacs{PACS numbers: 98.80.Cq\hfill Preprint PU-ICG-02/15, astro-ph/0208055v2}

\vskip2pc]

\section{Introduction}

It is now clear that the origin of structure in the Universe is a
primordial density perturbation, existing already when
cosmological scales start to enter the horizon. Observation is
{\em consistent} with the hypothesis that the density perturbation
is perfectly adiabatic, Gaussian and scale--independent, but
significant departures from this state of affairs is still allowed
by the data.  In particular it is not excluded that the adiabatic
density perturbation may be accompanied by a significant
isocurvature density perturbation \cite{Durrer,Amendola}.

Inflation provides a natural origin for the perturbation, since it
converts the vacuum fluctuation of each light free scalar field into a
classical scale--independent perturbation. One or more of these field
perturbations may cause the primordial density perturbation.

It is usually assumed that inflation involves a slowly-rolling
field, dubbed the inflaton, whose value determines the end of
inflation. The perturbation in the inflaton field cannot cause an
isocurvature perturbation, but does inevitably cause at some level
an adiabatic perturbation. The  usual  hypothesis \cite{LLbook} is
that the inflaton is solely responsible for the observed adiabatic
density perturbation. Under this `inflaton hypothesis' significant
non-Gaussianity is excluded in the usual one--field models
\cite{fnl,ks}.
In multi--field models, where there is a family
of possible inflaton trajectories curved in field space,
significant non-Gaussianity is possible \cite{daveetc,toni} but
apparently only at the expense of extreme  fine--tuning of the
initial condition that specifies the inflaton trajectory.
Any isocurvature density perturbation must be caused by the
perturbation of some non--inflaton field. Under the inflaton
hypothesis this means that the isocurvature density perturbation (if
present) depends on different physical parameters from the adiabatic
density perturbation. As a result, an isocurvature perturbation of
observable magnitude would require fine-tuning of the physical
parameters, or else some as-yet unforseen connection between them.

An alternative hypothesis
\cite{lm,LW,Moroi,MT2} is that the adiabatic density perturbation
originates from the perturbation in some `curvaton' field
different from the inflaton. In this scenario the adiabatic
density perturbation is generated
only after inflation, from an initial condition which corresponds
to a purely isocurvature perturbation
\cite{sylvia}.~\footnote{This conversion mechanism has been
considered also in the pre big bang scenario \cite{Enqvist,bggv}.
In this scenario though,  the required scale-invariant curvaton
field perturbations will be generated only if the curvaton has a
non-trivial coupling and for particular initial
conditions~\cite{CEW,LWC}.}
The object of the present 
paper is to explore the nature of the primordial density
perturbation under this hypothesis.
In the curvaton scenario, significant
non-Gaussianity may easily be present because the
curvaton density is proportional to the square of the
curvaton field. Also, the  curvaton density perturbation can lead,
after curvaton decay, to isocurvature perturbations in the densities
of the various components of the cosmic fluid. These, which we term
`residual' isocurvature components, are either fully correlated or
fully anti-correlated with the adiabatic density perturbation,
with a calculable and generally significant relative magnitude.

The paper is organized as follows. We deal in Section II with the
adiabatic perturbation and its possible non-Gaussianity.  In Section
III we formulate the description of isocurvature perturbations, in a
way which will allow us to analyse CDM, baryon and neutrino
perturbations in a unified manner.  In Section IV we calculate the
residual isocurvature perturbations of cold dark matter (CDM) and
baryons.  In Section V we give a general formalism for
describing the primordial neutrino isocurvature perturbation, taking
into account for the first time the crucial issue of lepton number. Then
we use it to calculate the residual isocurvature neutrino
perturbation.  Our conclusions are summarised in Section VI.

\section{The curvature perturbation}

\subsection{The curvature perturbation and the primordial density
perturbations}

{}From the viewpoint of observation the `primordial' epoch is the one
a few Hubble times before
 the smallest cosmological scale  approaches the horizon.
Taking that scale to enclose say $10^6$ solar masses, the
primordial epoch corresponds to temperature of order $10\keV$
which is  after nucleosynthesis.
Leaving aside the possibility of a
particle decaying after nucleosynthesis, the content of the Universe
at the primordial epoch is therefore rather well known.  There are photons,
practically massless neutrinos, baryons and (assuming it is already in
existence) cold dark matter, with the radiation dominating the energy
density.  The corresponding primordial density perturbation has an
adiabatic mode
\be
\frac14 \frac{\delta\rho_\gamma}{\rho_\gamma}=
\frac14 \frac{\delta\rho_\nu}{\rho_\nu}=
\frac13 \frac{\delta\rho_ B}{\rho_ B}=
\frac13 \frac{\delta\rho\sub{cdm}}{\rho\sub{cdm}}
\,,
\ee
which leaves the local ratio of number densities unperturbed.
Non-adiabatic or isocurvature modes are defined by
\bea
\S_B  &\equiv& \frac{\delta\rho_B}{\rho_B}-
\frac34\frac{\delta\rho_\gamma}{\rho_\gamma} \label{sb}\\
\S\sub{cdm} &\equiv& \frac{\delta\rho\sub{cdm}}{\rho\sub{cdm}}-
\frac34\frac{\delta\rho_\gamma}{\rho_\gamma}\label{scdm}\\
\S_\nu &\equiv& \frac34\frac{\delta\rho_\nu}{\rho_\nu}-
\frac34\frac{\delta\rho_\gamma}{\rho_\gamma}\label{snu}
\,.
\eea
In this section we focus on the adiabatic mode, which is known to be
the dominant one
responsible for structure formation, and return to
the possible isocurvature modes in the next section.

The primordial adiabatic density perturbation is associated with a
spatial curvature perturbation.
Following \cite{BST,WMLL} we define the curvature perturbation $\zeta$
on spatial slices of uniform density $\rho$ with the line element
\be
\diff\ell^2 = a^2 (1 + 2\zeta) \delta_{ij} \diff x^i \diff x^j \,.
\ee
The quantity $\zeta$ is related to the density perturbation,
$\delta\rho$, and curvature perturbation, $\psi$, on a generic
slicing (using the sign convention of Ref.\cite{MFB})
by the gauge-invariant formula\footnote
{In general a gauge corresponds to a definite
slicing and threading of spacetime, but in this
 paper only the former is relevant so that `gauge-invariant' can
be taken to mean `independent of the slicing'.}
\bea
\zeta
&=& -\psi-H\Delta t \label{curvtrans}\\
&=& -\psi - H\frac{\delta\rho}{\dot{\rho}}
\,.
\label{eqzeta}
\eea
where $\Delta t$ is the displacement of the generic
slicing from the uniform-density
slicing.
On super-horizon scales $\zeta$ is practically identical with the
curvature $\cal R$ defined on slices orthogonal to comoving
worldlines. The quantity $\zeta\approx \calr$ is useful on such scales
because it is time-independent~\cite{bardeen80,Lyth85,MS,WMLL,LLbook}
provided that the pressure perturbation is adiabatic, meaning that
$\delta P=c_s^2 \delta \rho$, where the adiabatic sound speed
$c_s^2=\dot{P}/\dot\rho$. This is guaranteed if there exists a
universal equation of state $P(\rho)$.
If the pressure perturbation is not adiabatic, then $\zeta\approx \calr$
changes according to the equation \cite{bardeen80,Lyth85,GBW}
\begin{equation}
 \label{zetadot}
 \dot\zeta = - {H \over \rho +P} \, \delta P_{\rm nad} \,,
\end{equation}
where $\delta P_{\rm nad}=\delta P - c_s^2\delta\rho$.

In the conventional inflaton model for the origin of structure
purely adiabatic perturbations are generated due to quantum
fluctuations in the single scalar field driving inflation. Thus
the curvature perturbation, $\zeta_*$, calculated shortly after
Hubble-exit ($k=aH$) determines the curvature perturbation until
that scale re-enters the Hubble scale during the subsequent
radiation or matter-dominated era.
By contrast, we are interested in a scenario where the curvature
perturbation generated on large-scales during inflation in the very early
universe is negligible.

In the rest of this section we describe the generation of the
large-scale curvature perturbation, $\zeta$, in the curvaton scenario,
amplifying the outline given in the original paper \cite{LW} (see also
\cite{moriond}). We deal with the simplest
version of the curvaton scenario, where the curvature perturbation is
caused exclusively by a single `curvaton' field distinct from the
inflaton field.

\subsection{Generating the curvaton field perturbation}

During inflation the Hubble parameter $H$ is assumed to be slowly
varying, $\epsilon_H\equiv \dot H/H^2 \ll 1$. The inflaton (if it
exists) is supposed to produce a negligible curvature perturbation.
Demanding that it is (say) less than $1\%$ of the observed value
implies \cite{LLbook} $V_*^\frac14< 2\times 10^{15}\GeV$, which in turn
implies \cite{ks} that the primordial gravitational waves will have no
detectable effect on the CMB anisotropy. Conversely, the detection of
an effect would rule out the curvaton hypothesis (an ``anti-smoking gun'')
\cite{WBMR,moriond}.

The curvaton field $\sigma$ is supposed to be practically free during
inflation, with small effective mass ($|V_{\sigma\sigma}| \ll
H^2$ where a subscript $\sigma$ denotes $\partial/\partial\sigma$).
It follows that on super-horizon scales there is a Gaussian
perturbation with an approximately scale-independent spectrum given by
\be
\calp_{\delta\sigma}^\frac12(k) \approx \frac{H_*}{2\pi}
\label{pinf}
\,,
\ee
where the * denotes the epoch of horizon exit $k=aH$. (The normalization
of the spectrum \cite{LLbook} is such that
$\calp_{\delta\sigma}^\frac12$  specifies the typical magnitude of
a spatial fluctuation in $\delta\sigma$ on a physical scale
$a/k$.) The spectral index specifying the slight scale-dependence
is given by \cite{LW,WBMR}
\be
 n-1\equiv \diff\ln(\calp_{\delta\sigma})/\diff\ln  k
  = 2\eta_{\sigma\sigma} -2\epsilon_H \,,
 \ee
where $\eta_{\sigma\sigma}\equiv V_{\sigma\sigma}/3H^2$.

After the smallest cosmological scale leaves the horizon, the
curvature perturbation remains negligible until after the curvaton
starts to oscillate.
As a result, the curvaton field
evolves in unperturbed spacetime. To follow the evolution it is
assumed that the curvaton field has no significant coupling to other
fields or, to be more precise, that the effect of any coupling can be
integrated out to give a possibly time-dependent effective potential
$V$.  With these assumptions, the curvaton field smoothed on the
smallest cosmological scale evolves along each comoving worldline
according to the unperturbed field equation
\be
\ddot \sigma + 3H\dot\sigma  + V_\sigma = 0
\,.
\ee
Making the first-order approximation
$\delta(V_\sigma(\bfx,t)) \approx V_{\sigma\sigma}(t)
 \delta\sigma(\bfx,t)$, its inhomogeneous perturbation satisfies
\be
\ddot {\delta\sigma} + 3H\dot{\delta\sigma} + V_{\sigma\sigma}\delta\sigma
=0
\,.
\ee
Assuming only that the evolution of $\delta\sigma$ is
linear, the Gaussianity and (slight) scale-dependence of the original
quantity are preserved. The fractional perturbation
$\delta\sigma(\bfx)/\sigma$ does not evolve at all if the potential is
either sufficiently flat (so that $\sigma$ and $\delta \sigma$ are
both constant) or quadratic (so that $\sigma$ and $\delta\sigma$
satisfy the same equation).

The field remains over-damped until
the Hubble parameter falls  below the  curvaton mass $m_\sigma$.
The curvaton field will then start to oscillate about its vacuum
value (taken to be $\sigma=0$) with an amplitude which decreases with
time. Even if the potential of the curvaton field is not quadratic,
after a few Hubble times we can make the approximation
$V\approx \frac12 m_\sigma^2 \sigma^2$.
The fractional field perturbation then has some
constant value,
\be
\(\frac{\delta\sigma}{\sigma}\)=
q\(\frac{\delta\sigma}{\sigma}\)_*
\,,
\ee
where the factor $q$ is time-independent because the oscillation is around
a quadratic minimum of the potential. In particular, if the effective
potential for the curvaton is quadratic (or sufficiently flat)
throughout  its evolution, $q=1$.
Using \eq{pinf}, the
spectrum of the fractional field perturbation at this stage is
\be
\calp^{1\over2}_{\delta\sigma/\sigma} = \frac q{2\pi} \frac{H_*}{\sigma_*}
\label{posc}
\,.
\ee

The energy density in the oscillating field is
\be
\rho_\sigma(\bfx,t) =
m_\sigma^2 \tilde\sigma^2(\bfx,t)
\,,
\label{rhosigma}
\ee
where $\tilde \sigma(\bfx,t)$ is the amplitude of the oscillation.
The perturbation in $\rho_\sigma$ depends on the curvaton field
perturbation through both a linear and a quadratic term. Assuming, for
the moment, that the linear term dominates (which we shall see is
demanded by the data) we have
\be
\frac{\delta\rho_\sigma}{\rho_\sigma} \approx 2\frac{\delta\sigma}
{\sigma} = 2q\(\frac{\delta\sigma}{\sigma}\)_*
\label{deltarhosigma}
\,.
\ee
We shall return to consider the possible contribution from the
quadratic term in section~\ref{nongauss}.

\subsection{Generating the  curvature perturbation}

So far we have reached the epoch just after the Hubble parameter falls
below the curvaton mass, and the curvaton oscillation starts. At this
stage, it is supposed that the dominant energy density comes from
radiation.  The curvaton, though, is supposed to be fairly long-lived,
while decaying comfortably before nucleosynthesis. So long as the
decay-rate is negligible ($\Gamma\ll H$), we have $\rho_\sigma \propto
a^{-3}$ and $\rho_r\propto a^{-4}$, leading to
$\rho_\sigma/\rho_r\propto a$. It is this increase in the relative
curvaton energy density which generates the curvature perturbation.

To analyse the generation of the curvature perturbation (and also,
later on, to discuss possible isocurvature perturbations produced by
the curvaton decay), it is convenient to consider the curvature
perturbations $\zeta_i$ corresponding to the separate components of
the energy density.  These are defined on slices of uniform $\rho_i$,
corresponding to the gauge-invariant definition
\begin{equation}
 \label{zetai}
\zeta_{i}
 \equiv - \psi - H\left(\frac{\delta\rho_{i}}{\dot{\rho_{i}}}\right)
\,.
\end{equation}
In particular, evaluating $\zeta_\sigma$ for the curvaton on
unperturbed ($\psi=0$) hypersurfaces when the curvaton starts to
oscillate, we have
\be
\zeta_\sigma = {1\over3} {\delta\rho_\sigma\over\rho_\sigma} =
{2\over3} q \left( {\delta\sigma\over\sigma}\right)_* \,.
\ee

The total curvature perturbation~(\ref{eqzeta}) can then
be written as the weighted sum
\begin{equation}
 \label{zetasum}
\zeta=(1-f)\zeta_r+f\zeta_\sigma
\end{equation}
where the relative contribution of the curvaton to the total
curvature is given by
\begin{equation}
\label{deff}
 f = \frac{3\rho_\sigma}{4\rho_r +
 3\rho_\sigma} \,.
\end{equation}
Thus the curvaton perturbation, $\zeta_\sigma$, can initially be
described as an isocurvature perturbation since
$f\to0$, i.e., $\rho_\sigma/\rho_r\to0$, in the early-time limit.

Until the effect of curvaton decay becomes significant, the curvaton
and radiation densities each satisfy their own energy conservation
equation $\dot\rho_i=-3H(\rho_i+P_i)$. In this regime,
each $\zeta_i$
is constant on super-horizon scales~\cite{WMLL}.
The evolution of $\zeta$ on these scales is due solely to the
change of $f$ in Eq.~(\ref{zetasum}), which yields
\begin{equation}
\label{dotf}
 \dot{\zeta}
 = \dot{f} \left(\zeta_{\sigma}-\zeta_r\right)
 = H f (1-f) \left(\zeta_{\sigma}-\zeta_r\right)
\end{equation}
It is intuitively easy to see how $\zeta$ changes with time as the
curvaton density grows relative to the radiation, $\dot{f}>0$.
It is straightforward to check that Eq.~({\ref{dotf}) is
consistent with Eq.~(\ref{zetadot}) where the non-adiabatic
pressure perturbation is given by 
%
\begin{equation}
\label{pnad}
 \delta P_{\rm nad}
 = {4 \rho_r \rho_\sigma \over 4\rho_\sigma + 3\rho_r}
 \left( \zeta_\sigma - \zeta_r \right) \,.
\end{equation}
An adiabatic perturbation
$\delta\rho_r/\dot\rho_r=\delta\rho_\sigma/\dot\rho_\sigma$
corresponds to the special case
\begin{equation}
\zeta=\zeta_r=\zeta_\sigma={\rm constant}\,.
\end{equation}

The curvaton scenario
corresponds to the case where the curvature perturbation in the
radiation produced at the end of inflation is negligible,
$\zeta_r\approx0$.
In the approximation of sudden decay \cite{LW}, $\zeta_r$ and
$\zeta_\sigma$ both remain constant up until decay and
the curvature perturbation at decay is therefore
\bea
\zeta &\approx&  f\sub{dec} \zeta_\sigma \label{fdec1} \\
&\approx& \frac{2}{3} f\sub{dec} q\(\frac{\delta\sigma}{\sigma}\)_*
\,,
\label{fdec}
\eea
where $f\sub{dec}$ is $f$ at the decay epoch,  conventionally  defined
in terms of the decay rate by $H\sub{dec}=\Gamma$.
We assume that after decay, the pre-existing radiation is either
insignificant, or else thermalizes with the decay products with the possible
exception of CDM. Since we are assuming that there is only one curvaton field,
this is sufficient to ensure that
$\zeta$ 
on cosmological scales will retain the same value until the primordial epoch.

Going beyond the sudden decay approximation, we define a number $r$
by
\bea
\zeta &=& r \zeta_\sigma \label{curvpred4}\\
&=&  r \frac13 \frac{\delta\rho_\sigma}{\rho_\sigma} \label{curvpred1}\\
&=& r q \frac{2}{3} \(\frac{\delta\sigma}{\sigma}\)_*
\label{curvpred3}
 \,,
\eea
where $\zeta$ is evaluated well after the epoch of
curvaton decay  and $\zeta_\sigma$ is evaluated well before this epoch.
In the limit where the curvaton completely dominates the energy density
before it decays, $r=1$. In other words, the sudden-decay approximation
becomes exact in this limit.
The reason is that in this case the curvaton and its
decay products constitute a single fluid with a definite $P(\rho)$,
so that they have 
 constant curvature perturbation which is equal
to $\zeta_\sigma$.
If the curvaton does not dominate, one has to resort to numerical
calculation of the coupled perturbation equations \cite{MW}, for which
one finds
\be
r\approx \(\frac{\rho_\sigma}{\rho} \)\sub{dec}
\label{rdef}
\,.
\ee

The prediction of the curvaton model for the spectrum of the curvature
perturbation is
\be
\calp_\zeta^\frac12 = \frac23 r \calp^\frac12_{\delta\sigma/\sigma}
\,,
\label{pred}
\ee
with $\calp^\frac12_{\delta\sigma/\sigma}$ given by \eq{posc}.
The COBE measurement of the CMB quadrupole anisotropy requires~\cite{LLbook}
\be
\calp_\zeta^\frac12({\rm COBE})
=4.8\times 10^{-5}
\label{cobe}
\,.
\ee
If the curvaton dominates
the energy density before it decays ($r=1$) this implies the following
amplitude for the perturbations of the curvaton field
\be
\calp^\frac12_{\delta\sigma/\sigma}
= 7.2\times 10^{-5}
\label{r1pred}
\,.
\ee
In other words, the typical field (and density) perturbation on
cosmological scales must be of order $10^{-4}$ in that case.  If the
curvaton does not dominate, we need at least $r\gsim 10^{-4}$ to get a
spectrum of the correct magnitude (since the typical density
perturbation can be at most of order 1). We shall now see how the
observational bound on non-Gaussianity actually requires a much higher
value.

\subsection{Possible non-Gaussianity of the curvature perturbation}

\label{nongauss}

{}From \eqs{rhosigma}{curvpred1}, the curvature perturbation depends
on the curvaton field perturbation through both linear and a quadratic
terms. So far we evaluated only the linear contribution, which gives a
Gaussian contribution to the curvature perturbation. However, if the
quadratic term in the density perturbation of the curvaton field is
not negligible, the total curvature perturbation will have a
non-gaussian ($\chi^2$) component.

The relative magnitude of the quadratic term is
conventionally specified \cite{ks} by a number $f\sub{NL}$
(NL meaning `non-linear'), which nominally  determines
 the non-Gaussian contribution to
the Bardeen potential according to the formula
\be
\Phi = \Phi\sub{gauss}  + f\sub{NL}
 \Phi^2\sub{gauss}
\label{fnldef}
\ee
The connection with the Bardeen potential is only nominal, because
the relation  between $\Phi$ and $\zeta$  is taken
\cite{ks} to be
\be
\Phi = -\frac35 \zeta = -\frac{r}{5} {\delta\rho_\sigma\over\rho_\sigma} \,.
\label{Phizeta}
\ee
This relation
is actually valid only on super-horizon scales after matter domination,
the  correct relation  at the primordial epoch being more complicated
and  involving
the relative neutrino density \cite{LLbook}.

Using \eq{rhosigma} we have
\bea
\frac{\delta\rho_\sigma}{\rho_\sigma} &=&
 2\frac{\delta\sigma}{\sigma} +
\frac{(\delta\sigma)^2}{\sigma^2}
\,.
\eea
Thus, using \eqs{fnldef}{Phizeta}, the prediction of the
curvaton hypothesis is
\be
f\sub{NL} = \frac{5}{4r}
\label{fnlpred} \,.
\ee

At this point we have to remember that first-order cosmological
perturbation theory is being assumed. Second-order metric
perturbations will also give a correction $\Phi^{(2)}(\bfx)$,
presumably with typical magnitude
\be
|\Phi^{(2)}(\bfx)|\sim {\Phi\sub{gauss}^2}
\,.
\label{phi2}
\ee
It follows that the estimate \eq{fnlpred} of $f\sub{NL}$ will be valid
only in the regime $f\sub{NL}\gg 1$ (and that smaller values of
$f\sub{NL}$ cannot even be defined, unless $\Phi^{(2)}(\bfx)$ is
itself the square of a Gaussian quantity). In other words, the
validity of the estimate \eq{fnlpred} requires that the curvaton
contributes only a small fraction of the energy density just before it
decays. In the opposite case that the curvaton dominates the density
before it decays, the non-Gaussianity calculated according to linear
theory is lost in the ``noise'' of the unknown second-order correction to
cosmological perturbation theory.

Now we compare the prediction with present and future data on the CMB
anisotropy.  A recent analysis of the COBE data \cite{data} yields
$|f\sub{NL}| \lsim
2\times 10^{3}$, implying the constraint $r\gsim 6\times 10^{-4}$.
{}From \eqs{pred}{cobe} we see that this bound is equivalent to
$\calp^\frac12_{\delta\sigma/\sigma}\lsim 0.1$.
Starting 
in 2003, data from the MAP satellite~\cite{MAP} will either detect
non-Gaussianity or give $|f\sub{NL}| \lsim 20$ \cite{ks} corresponding
to $r\gsim 0.06$. If non-Gaussianity is detected by MAP it will clearly
be above the noise of the second-order correction.  Looking further
ahead, the PLANCK satellite will either detect non-Gaussianity or give
$|f\sub{NL}| \lsim 5 $ \cite{ks} corresponding to $r\gsim 0.2$.
Apparent non-Gaussianity at the bottom end of this range would however
have to be checked against the second-order order correction to $\Phi$
to make sure that the correction is negligible.

We have seen that the curvaton hypothesis can easily give significant
non-Gaussianity. The reason is that the predicted curvature
perturbation is proportional to  the perturbation of the curvaton
density, a quantity which depends on the {\em square} of the curvaton
field. The perturbation in the curvaton field is supposed to be
Gaussian because it is supposed to have negligible interaction, making
its Fourier components uncorrelated which is the definition of
Gaussianity. But the corresponding perturbation in the curvaton
density is a linear combination of the curvaton field perturbation and
its square. {}From a theoretical viewpoint the square could even
dominate \cite{lythaxion,lm}, though as we have seen this is ruled out
by the data.

This is in sharp contrast with the situation for the inflaton
hypothesis, in which the curvature perturbation is purely linear in
the inflaton field perturbation. The inflaton can usually be treated
as a practically free field, making its perturbations and the
curvature practically Gaussian. In particular, for the usual
single-field models, where there is an essentially unique inflaton
trajectory, the self-interaction of the inflaton field is the only
relevant one and is kept small by the flatness conditions $\epsilon,
\eta\ll 1$.  As a result single-field models give \cite{fnl}
$|f\sub{NL}|=|2\epsilon - 2\eta| \lsim 0.1$ which is lost in the noise
from the second-order correction to cosmological perturbation 
theory.\footnote{The bound $0.1$ comes from the observational bound
on the spectral index in the inflaton scenario, $n=2\eta-6\epsilon$.}
In two-field models, where there is a family of possible inflaton
trajectories curved in field space, the inflaton field perturbation at
the end of inflation will be a linear combination \cite{gwbm}
\be
\delta\phi\sub{end}
 \propto
 \cos\theta \delta\phi_* + \sin\theta \delta\chi_*
\,,
\ee
where $\phi_*$ is the inflaton field at the time of horizon exit
and $\chi_*$ is the field orthogonal to it. By definition the slope
of the potential vanishes in the $\chi$ direction but the higher derivatives
might in principle be present corresponding to a self-interaction big enough
to generate significant non-Gaussianity. In particular
there might be
a cubic term $M\chi^3$ in the potential  \cite{bmr}
(with the unperturbed $\chi_*=0$)
or something more complicated \cite{daveetc}. This setup might generate
significant non-Gaussianity, but it obviously requires a special
choice of the inflaton trajectory. Even when such a choice is made initially,
it might be destabilized by the quantum fluctuation before cosmological
scales start to leave the horizon.

We conclude that a detection of non-Gaussianity by MAP would strongly
suggest that the primordial curvature perturbation is produced by
a curvaton field. Such a detection would imply that the density of the
curvaton before it decays reaches no more than $6\%$ of the total.
In the opposite case that the curvaton  dominates before it decays,
it gives a practically Gaussian curvature perturbation, just
as in the inflaton scenario.


\section{Isocurvature perturbations}

Thus far we have concentrated upon how the curvaton introduces a
large-scale curvature perturbation. Now we consider the {\em
isocurvature} perturbations that may be imprinted in different
particle species after the curvaton decay.

We  adopt the `separate universe' viewpoint, implicit in
practically all discussions of   perturbations
on super-horizon scales.
At each epoch, it  is assumed that each comoving region with size
much bigger than the  Hubble distance looks locally like some unperturbed
(Robertson-Walker) universe.
If these `separate universes' are all identical,  the cosmological
perturbations  are said to be adiabatic. Perturbations in matter
fields then vanish on slices of uniform energy density, and the
curvature perturbation $\zeta$ is the only thing that needs to be
specified to determine the evolution of
the perturbations after horizon entry.

If the `separate universes' are not identical there are
isocurvature perturbations, so-called
because they evolve independently of the curvature perturbation on
large scales which can therefore  be taken to vanish when
considering them.
One way of specifying a generic isocurvature perturbation,
$\delta x$, is to give its value on uniform-density slices, related to
its value on a different slicing by the gauge-invariant equation
\be
H\(\frac{\delta x}{\dot x} \)_{\delta\rho=0}
=H\( \frac{\delta x}{\dot x} - \frac{\delta\rho}{\dot\rho} \)
\label{gud}
\,.
\ee
For a set of fluids with energy density $\rho_i$,  the isocurvature
perturbations are instead conventionally defined by the
 gauge-invariant quantities
\be
 \S_{ij} = - 3H\(\frac{\delta\rho_i}{\dot\rho_i}-
\frac{\delta\rho_j}{\dot\rho_j}\)
 \label{sij1}
\,.
\ee
In terms of the individual curvature perturbations defined in
Eq.~(\ref{zetai}) this becomes
\be
\S_{ij} = 3\( \zeta_i - \zeta_j\)
 \label{sij2}
\ee
For  fluids which are `decoupled'
(in the sense that there is no energy
transfer) and which have  a definite equation of state $P_i(\rho_i)$,
the $\zeta_i$'s are constant on superhorizon scales~\cite{WMLL} and
hence so
are the isocurvature perturbations (\ref{sij2}). In this language, the
curvaton density perturbation before it decays corresponds to a
constant isocurvature perturbation $S_{\sigma r}\equiv
3(\zeta_\sigma-\zeta_r)$ which in the curvaton scenario ($\zeta_r=0$)
reduces to $\S_{\sigma r}\equiv3\zeta_\sigma$.

At the `primordial' epoch, before the smallest cosmological scale
approaches the horizon,  there are four `decoupled' fluids namely
the CDM, baryons, photons and  neutrinos (taken to be massless).
The  conventional definitions \eqst{sb}{snu} of the three
isocurvature perturbations correspond to
\bea
\label{SB}
\S_B &=& 3\(\zeta_B -\zeta_\gamma\) \\
\label{Scdm}
\S\sub{cdm} &=& 3\(\zeta\sub{cdm} - \zeta_\gamma \)\\
\label{Snu}
\S_\nu &=& 3\( \zeta_\nu - \zeta_\gamma\)
\,,
\eea

In the standard single-inflaton scenario it is impossible to
introduce isocurvature perturbations on large scales at the
primordial epoch from the purely adiabatic perturbations existing
after the end of inflation~\cite{WMLL}.  But in the curvaton
scenario the non-adiabatic nature of the curvaton perturbation
after inflation ($\S_{\sigma r}\neq 0$) means that it is possible
for the curvaton to leave isocurvature perturbations after the
curvaton decays even on super-horizon scales, which we term
`residual' isocurvature perturbations.


In addition to the energy densities we shall need to consider
`number' densities
$n\sub{cdm}$, $n_B$,  $n_L$ and  $n_{Li}$.  They are
defined, respectively, as (i) the number density of CDM particles,
(ii)
the density of baryon number $B$,  (iii) the density of lepton number
$L$ and (iv)
 the densities of the three individual lepton
numbers $L_i$ ($i=$ e, $\mu$ or $\tau$).
 By the `primordial' era
any significant lepton number will be carried entirely by neutrinos,
making $n_{Li}$ the difference between neutrino and anti-neutrino
number densities.
In the unperturbed Universe each number density is
proportional to $a^{-3}$
so long as the corresponding quantity is conserved in a comoving
volume.

For each of these number densities, it
 will be  useful to consider  the curvature perturbations
$\tilde \zeta_i$ on slices of uniform $n_i$,
\bea
 \label{altzetai}
\tilde\zeta_i &\equiv& -\psi - H {\delta n_i \over \dot{n}_i} \\
&=& -\psi+ \frac13 {\delta n_i \over n_i} \,.
\eea
In a homogeneous universe the conserved number density $n_i$ obeys the
evolution equation
\be
\dot{n}_i + 3H n_i = 0 \,.
 \label{dn}
\ee Allowing for large-scale perturbations about the strictly
homogeneous background, we obtain the local evolution equation for
the first-order perturbation
 \be
 \dot{\delta n_i} + 3H\delta n_i - 3n_i\dot\psi = 0 \,,
 \label{ddeltan}
\ee
assuming all spatial gradients (e.g., divergence of the particles
velocity field) are negligible on large scales.
Combining equations (\ref{altzetai}--\ref{ddeltan}) we obtain the
evolution equation for the curvature perturbation
\be
\dot{\tilde\zeta}_i = 0 \,.
 \label{altdzetai}
\ee
Hence we see that $\tilde\zeta_i$ defined in Eq.~(\ref{altzetai}) remains
constant on large-scales so long as the particle number $n_i$ is
conserved.

Equations (\ref{altzetai}) and (\ref{altdzetai}) are equivalent to the
statement that the perturbation in the fractional number density
($\delta n_i/n_i$) is conserved on flat ($\psi=0$) spatial
hypersurfaces.  The perturbations in the number densities are best
defined on the flat slicing, because {\em on this slicing the
  expansion rate with respect to coordinate time is unperturbed}
  \cite{WMLL}{\em, leading to the constancy of the fractional
  perturbations}.

This is a significant extension to the case of interacting fluids of
the result found in Ref.\cite{WMLL} for the constancy of $\zeta_i$ for
non-interacting fluids on large-scales. It will prove a powerful tool
to describe the possible generation of isocurvature perturbations
after curvaton decay in what follows.

\section{Residual  isocurvature matter  perturbations}

\subsection{Residual isocurvature  CDM perturbation}

Without making any assumption about the nature of the CDM, we take the
epoch of its creation as the one after which its particle number
$n\sub{cdm}$ is conserved. There are several candidates for the CDM
particle, such as the axion, the lightest supersymmetric particle
(LSP), a stable massive particle (wimpzilla) or primordial black
holes. For axions the epoch of creation corresponds to the temperature
$\sim 1\GeV$ at which the effective axion mass rises above the Hubble
parameter, and for the conventional LSP it corresponds to the
freeze-out temperature roughly of order $10\GeV$. For very massive
particles (wimpzillas) \cite{lrs,ckr}
or black holes \cite{rsg,glw}
the epoch of creation may be the
end of slow-roll inflation corresponding to an
energy scale $\rho^\frac14$ up to $10^{16}\GeV$. In the case of
wimpzillas it may instead \cite{ewanhui} be the epoch of
thermal inflation
 corresponding to perhaps $\rho^\frac14 \sim
10^6\GeV$. The CDM might also be created as an out-of-equilibrium
decay product of the inflaton, the curvaton or some other particle.

After CDM creation, conservation of the CDM particle number ensures that
 $\tilde\zeta\sub{cdm}$ defined by~Eq.~(\ref{altzetai}) is constant
on super-horizon scales .
When  the  CDM becomes non-relativistic with constant mass,
 $\tilde\zeta\sub{cdm}$
reduces to $\zeta\sub{cdm}$ defined by Eq.~(\ref{zetai}).
In most of the examples mentioned above, this  occurs at the epoch of
creation. It could happen though that that CDM created from 
out-of-equilibrium decay is initially relativistic. Also, the axion
mass (induced by the QCD instanton)
increases after the epoch of creation, becoming constant only 
when the temperature falls to  $100\MeV$. 
In any case, $\tilde \zeta\sub{cdm}$ will certainly have reduced to
$\zeta\sub{cdm}$ by the primordial epoch.

To evaluate the primordial isocurvature perturbation $\S\sub{cdm}$, we
assume that the primordial $\zeta_\gamma=\zeta$ corresponding to a
zero or small neutrino isocurvature perturbation
$\zeta_\nu-\zeta_\gamma$, postponing until Section~\ref{isonu} the
possibility of a significant neutrino isocurvature perturbation. 
We can then re-write \eq{Scdm} as
\be
 \S\sub{cdm} = 3\( \tilde \zeta\sub{cdm} - \zeta \) \,.  
\ee

As we have mentioned, several different mechanism have been considered
for creating the CDM. In most cases, the creation mechanism does not
involve any quantity with an isocurvature perturbation, which means
that at creation the CDM has no isocurvature perturbation. To put it
differently, the `separate universes' are in these cases identical at
the epoch of CDM creation, which means that the number density
$n\sub{cdm}$ at creation depends only on the local energy density.

The most usual way of obtaining CDM which has an isocurvature
perturbation at creation is to suppose that it consists of axions,
whose production in each region of space corresponds to the onset of
the axion field oscillation in that region. If the axion field has an
isocurvature perturbation, a CDM isocurvature perturbation will then
be produced at the time of creation.
However, the magnitude of the isocurvature CDM perturbation produced
in this way is unrelated to that of the curvature perturbation. In
other words, there is no reason why CDM originating from an
oscillation of the axion field should have a significant isocurvature
density at the time of its creation.  The same is true for all of the
other CDM production mechanisms that have been considered so far.

In what follows we will assume that the CDM number density at the
epoch of creation depends only on the local energy density.  We expect
this to be valid in the absence of any non-adiabatic pressure
perturbation, i.e., where the local density also determines the local
pressure. 
The only exception will be the case that the CDM is produced directly
by the curvaton decay, which will require separate treatment.

In the curvaton scenario where the CDM is created after the curvaton
has decayed (or in the inflaton scenario) the CDM isocurvature
perturbation will be zero at the `primordial' epoch.  The basic reason
is that the `separate Universes' in these cases are identical. To
proceed more formally, the assumption that there is initially no
isocurvature density perturbation means that the CDM number density at
creation is uniform ($\delta n\sub{cdm}=0$) on slices of uniform
density (where $\psi=-\zeta$).  Hence we may evaluate the
gauge-invariant expression (\ref{altzetai}) on a surface of uniform
density to obtain $\tilde\zeta\sub{cdm} =\zeta$, with both sides
constant. Going forward to the primordial epoch, when the CDM energy
is conserved so that $\zeta\sub{cdm}=\tilde\zeta\sub{cdm}$ we find
that the primordial CDM isocurvature perturbation, $\S\sub{cdm}$ given
by Eq.~(\ref{Scdm}), vanishes.

The situation is dramatically different if the CDM is created before
the curvaton decays, or if it is created by the curvaton decay itself.
In these cases, the process of curvaton decay creates a `residual'
isocurvature CDM perturbation whose properties are closely tied to the
curvature perturbation.

Consider the case that the CDM is created well before the curvaton
decays 
and well before the curvaton comes to dominate the energy density ($f\ll1$
in \eq{deff}). 
In this regime the Universe is practically unperturbed and therefore
$\tilde\zeta\sub{cdm}$ is practically zero.
At  the primordial epoch this gives
the residual CDM isocurvature perturbation in \eq{Scdm} as
\be
 \S\sub{cdm} = - 3\zeta
\label{Scdmzeta}
 \,.
 \ee
In the notation of Ref.~\cite{Amendola} this corresponds to
a maximum correlation between curvature and CDM-isocurvature
perturbations. The power spectra of the two perturbations have the
same spectral index and the isocurvature perturbation has an
amplitude three times larger than the adiabatic one.  Such a large
correlated perturbation is ruled out by current observations which
require\cite{Amendola,luca}
\be
\left| {\S\sub{cdm} \over \zeta} \right| < 1.5 \ {\rm at}\ 95\% {\rm
  c.l.} \,.
 \label{cdmcon}
\ee

Our conclusion  is that in the curvaton scenario, 
{\em CDM cannot be created before the curvaton decays and while the
  total curvature perturbation remains small, $\zeta\ll
  r\zeta_\sigma$}. In particular CDM creation just after inflation
ends, such as might occur in the case of wimpzillas or black holes, is
incompatible with the curvaton scenario.
%

If the CDM is created before the curvaton decays, but when the
curvaton density has become non-negligible, $f\sim1$ in \eq{deff},
there will be a significant non-adiabatic pressure perturbation,
$\delta P\sub{nad}$ given by \eq{pnad}, at the epoch of creation and
we can no longer assume that the initial CDM number density is
unperturbed on uniform density slices. Instead the actual number
density at creation will depend on the mechanism by which the CDM is
created, and we leave a detailed investigation of the different cases
for future work.


Finally, we consider the case that the CDM is created by the decay of
the curvaton itself. The epoch of CDM creation then corresponds to the
epoch when the curvaton decay is complete. The resulting local CDM
density is then a fixed multiple of the curvaton number density well
before decay. The fractional perturbations are thus equal and hence
\be
\tilde\zeta\sub{cdm} = \zeta_\sigma \,.
\ee
Using the definition of $r$ in \eq{curvpred4},
this gives at the primordial epoch
\be
\S\sub{cdm} =
3\( \frac{1-r}r \) \zeta
\,.
\ee
This is negligible if the curvaton comes to dominate before it decays
($r=1$), the physical reason being that the curvaton perturbation
becomes an adiabatic curvature perturbation, so it cannot leave
behind any residual isocurvature perturbation.
But if the curvaton decays before it dominates,
$r\ll1$, there will be large isocurvature
perturbations in the CDM, incompatible with existing observational
constraints. Note that in the notation of Ref.\cite{Amendola} the
curvature and isocurvature perturbations are anti-correlated, in which
case the observational limit becomes
\be
\left| {\S\sub{cdm} \over \zeta} \right| < 0.2 \ {\rm at}\ 95\% {\rm
  c.l.} \,,
 \label{anticdmcon}
\ee
much stronger than that for correlated isocurvature perturbations.
Observational limits on the amplitude of the anti-correlated
isocurvature perturbations thus require the curvaton to effectively
dominate, $r>0.9$, if the CDM is created by the curvaton decay.

\subsection{Residual baryon isocurvature perturbation}

Baryon number $B$ may be created directly, in which case we take
the epoch of baryogenesis  to be the one after which $B$ is
conserved. Alternatively $B$ may be created through the conversion
of $L$ at the electroweak transition. In the latter case $B-L$ is
conserved, and for the present purpose we may take the epoch of
baryogenesis
 as the one when $B-L$ is created.  Depending on
what type of mechanism operates, the epoch of baryon creation in
this sense may be anywhere from the end of inflation to the
electroweak transition.  To keep the notation simple we take the
relevant quantity to be $B$ from now on, with the understanding
that $B$ is to be replaced by $B-L$ if that is the relevant
quantity.

Until the QCD transition at $T\sim 100\MeV$, $B$ represents the
difference between the abundances of the typically relativistic
particles and anti-particles carrying baryon number, and cannot
usefully be associated with an energy density. In this situation we
need to use $\tilde\zeta_B$ the curvature perturbation on slices of
uniform $n_B$ (given in Eq.~(\ref{altzetai})).  This quantity is
constant on large scales at all times after baryon creation, and it
reduces to $\zeta_B$ (Eq.~(\ref{zetai})) after the QCD transition when
baryon number is carried by non-relativistic nucleons and nuclei.

The theoretical situation for the baryon isocurvature perturbation is
similar to the one for CDM.  We make the reasonable assumption that
the baryon number density at the epoch of baryogenesis depends only on
the local energy density, unless baryogenesis comes the curvaton decay
itself.  If baryogenesis occurs after the curvaton decays, there will
be no primordial baryon isocurvature perturbation.  If instead baryon
number is generated well before curvaton decay 
and well before the curvaton dominates,
%
there will be a large residual baryon
isocurvature perturbation,
\be
 \S_B = -3 \zeta
 \label{SBzeta}
\,.
\ee
In contrast with  the CDM case,  such a perturbation
is still marginally compatible with current observational
data~\cite{Amendola}. 
The effect on the CMB angular power spectrum of a baryon-isocurvature
perturbation is essentially the same as that of a CDM-isocurvature
perturbation but the size of the effect is diminished
by a factor $\Omega_B/\Omega\sub{cdm}$ due to the smaller density of
baryonic matter \cite{BMT}, so the constraint in Eq.~(\ref{cdmcon})
becomes
\be
\left| {\S_B \over \zeta} \right|
 < 1.5 \({\Omega\sub{cdm}\over\Omega_B}\) \ {\rm at}\ 95\% {\rm
  c.l.} \,.
 \label{Bcon}
\ee
An isocurvature perturbation of the form given in Eq.~(\ref{SBzeta})
will either be ruled out or observed in the near future, providing in
the latter case a smoking gun for the curvaton scenario.


The final possibility is that baryon number is produced by the
out-of-equilibrium decay of the curvaton itself, as is the case if we
identify the curvaton with the right-handed sneutrino of ref.\
\cite{yanagida}.  Then we have at the primordial epoch,
in the sudden decay approximation,
\be
\S_B =
3\(
{1-r\over r} \) \zeta \,.
\ee
Current observational limits on the amplitude of such an
anti-correlated baryon-isocurvature mode
require that the curvaton must dominate the density at decay, $r>0.6$,
if the curvaton itself is responsible for baryogenesis.

\section{Neutrino isocurvature perturbations}

\subsection{The general formalism}
\label{isonu}

In this section we discuss the possibility of a neutrino
isocurvature density perturbation. As far as we are aware our
treatment is the first one taking into account the crucial issue of
lepton number.

The era of thermal equilibrium for neutrinos ends just before
nucleosynthesis, which  almost certainly means that this era
determines the abundances of neutrinos and anti-neutrinos at the
later `primordial' epoch. (We are not in this paper considering
the possibility that the curvaton decays after nucleosynthesis,
and for the present purpose we discount too the possibility that
it decays between neutrino decoupling and nucleosynthesis.) Taking
that for granted, there is  no neutrino isocurvature
if the lepton number at decoupling is
negligible, because the primordial abundance of the neutrinos is
then determined entirely by the photon temperature~\cite{BMT}.

We therefore consider the case of non-zero lepton number density
$n_{Li}$, 
with $i=$e, $\mu$ or
$\tau$.
 While the neutrinos are effectively massless and in
equilibrium, with temperature $T_\nu$ this corresponds to
occupation numbers
\be
 f_i(E)=[\exp(E/T_\nu\mp \xi_i) +1]^{-1}
\,,
\label{occnum}
\ee
for neutrinos and anti-neutrinos with energy $E$. This gives the
following expressions for the {\em total} energy density $\rho_i$
of neutrinos and anti-neutrinos, and the  lepton number
density $n_{Li}$ equal to the {\em difference} between the number
densities of neutrino and
anti-neutrinos \cite{dolgov};
\bea
 \frac{\rho_i}{\rho_\gamma} &=&
  \frac78 \(\frac{T_\nu}{T_\gamma}\)^4 A_i
   \label{fracrhoi}
\\
 \frac{n_{Li}}{n_\gamma}
  &=& 2.15 \(\frac{T_\nu}{T_\gamma}\)^3 B_i
   \label{fracni}
\,.
\eea
where
\bea
 A_i &=&
 \[ 1 + \frac{30}7
 \(\frac{\xi_i}\pi \)^2 + \frac{15}7  \(\frac{\xi_i}\pi \)^4
 \]\,,\\
 B_i &=& \[ \(\frac{\xi_i}\pi \) + \(\frac{\xi_i}\pi \)^3 \] \,,
\eea
 and
\bea
 \rho_\gamma &=& \frac{\pi^2}{15} T_\gamma^4 \,,\\
 n_\gamma &=& \frac{2.40}{\pi^2} T_\gamma^3 \,.
\eea

In the usual case that the lepton asymmetry is negligible,
these expressions start to become valid at reheating, and hold with
$T_\nu=T_\gamma$ until
 positron annihilation, after which
\be
T_\nu = (4/11)^\frac13  T_\gamma
\label{tnugamma}
\,.
\ee
In the case of significant lepton asymmetry that we are considering,
there is significant neutrino mixing and the expressions become
valid only when $T_\nu=T_\gamma$ falls to a few MeV.
The subsequent evolution is more complicated than in the usual
case so that the thermal
 distribution \eq{occnum} is not precisely maintained
\cite{dolgov}, but following the usual practice we make the approximation
that it is maintained so that \eqst{occnum}{tnugamma} are all valid
after positron annihilation.  {}From now on we take the expressions to
refer to that era. The asymmetry parameters $\xi_i$ are then constant
since the neutrinos are travelling freely leading to $n_i\propto T_\nu^3
\propto 1/a^3$.

Big-bang nucleosynthesis (BBN) and large-scale structure (LSS)
 constrain  $\xi_e$ and $|\xi_\mu|^2+|\xi_\tau|^2$.
If the favoured large mixing angle (LMA) solution to the solar neutrino
problem is correct, neutrino oscillations ensure that
the $\xi_i$ have  a common
value $\xi$ \cite{mixing} . In that case, the BBN/LSS constraint is
 \cite{Orito} $-0.01 < \xi <
0.07$. With this constraint, {\em unperturbed} lepton number
densities are almost certainly too small to have any observable
effect on the CMB anisotropy or large-scale structure (LSS). We
shall see, though, that this need not be the case for the residual
isocurvature perturbation.

Final  confirmation of the LMA is expected in a few months from
the Kamland reactor experiment \cite{kamland}, but we shall
nevertheless allow independent asymmetry parameters in our
analysis. In that case BBN and CMB/LSS together give constraints
on $|\xi_\mu|=|\xi_\tau|$ (assumed equal for convenience) and
$\xi_e$ which are much weaker, namely $|\xi_{\mu,\tau}|\leq2.5$
and $|\xi_e|\leq0.30$~\cite{Orito}. Note that these limits were
obtained assuming a purely adiabatic primordial perturbation
spectrum and hence serve only as a rough guide in the case of a
correlated curvature and neutrino isocurvature perturbations.

For each individual neutrino species we can calculate the
curvature $\zeta_i$  on uniform-$\rho_i$ hypersurfaces from
Eqs.~(\ref{zetai}) and~(\ref{fracrhoi})
\begin{equation}
 \zeta_i-\zeta_\gamma
= \({\delta T_\nu \over T_\nu}  - {\delta T_\gamma \over
  T_\gamma} \) + {1\over4}{A_i^\prime \over
  A_i} {\delta\xi_i\over\pi}
\label{nuzetai1}
\,,
\end{equation}
and the   curvature $\tilde \zeta_i$
on uniform-$n_{Li}$ hypersurfaces
 given from
Eqs.~(\ref{altzetai}) and~(\ref{fracni}) as
\begin{equation}
 \tilde\zeta_i -\zeta_\gamma =\( {\delta T_\nu \over T_\nu}
 - {\delta T_\gamma \over  T_\gamma} \)+
  {1\over3}{B_i^\prime \over B_i} {\delta\xi_i\over\pi} \,,
 \label{nuzetai}
\end{equation}
where $A_i'=(60/7)B_i$ and $B_i^\prime=1+3(\xi_i/\pi)^2$.
Only when $\delta\xi_i=0$ do these two quantities  coincide.

As we have noted, a neutrino isocurvature perturbation due to the
first term in \eqs{nuzetai1}{nuzetai}, representing a perturbation in
the neutrino to photon temperature ratio, is extremely unlikely since
it would have to be generated during the extremely short era between
neutrino decoupling and nucleosynthesis. We here take the first term
to be zero.

Following the path that we trod for the CDM and baryon number
perturbations, we define the isocurvature perturbations in the
number densities of the three lepton numbers,
\bea \tilde\S_i &=&
\frac{\delta n_{Li}}{n_i}-
\frac{\delta n_\gamma}{n_\gamma}  \\
&=& 3(\tilde \zeta_i- \zeta_\gamma )
\label{sidef}
\,.
\eea
They determine the perturbations in the asymmetry parameters,
\be
 {\delta\xi_i\over\pi} =  {B_i\over B_i'} \tilde\S_i
\approx \frac{\xi_i}\pi \tilde \S_i
\label{deltaxi}
\,,
\ee
where the final equality is valid in the regime  $(\xi_i/\pi)^2\ll 1$.
These  expressions refer to the primordial era, and are valid in the early
Universe only back to the epoch  of positron annihilation at $T\sim \MeV$.
While they are valid,  the lepton numbers $L_i$ are conserved and
the curvature perturbation $\tilde \zeta_i$ on slices of uniform $n_{Li}$
is constant.
At early times though,
corresponding to $T$ bigger than a few MeV, neutrino mixing
becomes significant and the individual lepton numbers are not defined.
Instead there is only the total lepton number density and its perturbation,
\bea
n_L &=& \sum n_{Li}\\
\tilde\S_L &=&
 \frac{\delta n_L}{n_L}-
\frac{\delta n_\gamma}{n_\gamma}
\\
&=& 3(\tilde \zeta_L- \zeta_\gamma )
 \,.
\eea
The total lepton number $L$ is well-defined in the early Universe
after the epoch of lepton number creation, which we take to be the
one after which $L$ is conserved. While $L$ is conserved, the
curvature perturbation $\tilde\zeta_L$ on slices of uniform $n_L$
is constant.

The three primordial quantities $n_i$ and their perturbations
$\S_i$ are not in general determined by $n_L$ and $\S_L$. They are
however determined if the LMA solution to the solar neutrino
problem is correct because the three quantities are then equal
\bea
 n_{Li} &=& {1\over3} n_L \\
 \tilde\S_i &=& \tilde\S_L
\eea

If one or more neutrino masses are big enough to give a significant
amount of dark matter, it will be necessary to insert the
$\delta\xi_i$ into the initial occupation number \eq{occnum},
propagating this initial condition forward with evolution equations.
In that case the initial condition requires a knowledge of the three
individual isocurvature perturbations $\S_i$. Here we make instead the
opposite approximation of massless neutrinos. Then, the only required
initial condition is the isocurvature perturbation in the total
neutrino energy density which is specified by the derived quantity
$\S_\nu$, defined by \eq{snu}. This quantity may be calculated using
\be
 \zeta_\nu = \frac{\sum \rho_i \zeta_i}{\sum \rho_i}
 \,,
\ee
taking $\rho_i$ from \eq{fracrhoi} and $\zeta_i$ from
\eq{nuzetai1}. The result is 
\bea
 S_\nu &=& {45\over7} \frac{\sum B_i (\delta\xi_i/\pi)}{\sum
 A_i} \,,\\
 &=&
  {45\over7} \frac{\sum  (B_i^2/B_i') \tilde\S_i}{\sum A_i}
  \approx
    {15\over7} \sum \left({\xi_i\over\pi}\right)^2 \tilde\S_i
 \label{finalSnu}
\,,
 \eea
where the final approximation is valid for $(\xi_i/\pi)^2\ll 1$.

As far as we are aware, this is the first time that expressions for
the neutrino
isocurvature perturbation have been given in the most realistic case
where it is determined by the (perturbed) lepton asymmetry.
The effect of the neutrino isocurvature perturbations has
never been tested against observational data while including
non-zero chemical potential.

\subsection{Induced matter isocurvature perturbations}

Now we address an issue that is relevant for calculating the
amplitude of CDM and baryon isocurvature perturbations in the
presence of a neutrino isocurvature perturbation. In the presence
of the latter, the equality $\zeta_\gamma=\zeta_\nu=\zeta$ breaks
down, and instead we have
\be
 \zeta  = (1-R_\nu)\zeta_\gamma + R_\nu \zeta_\nu
  = \zeta_\gamma + {R_\nu\over3} \S_\nu \,,
 \label{zetazetanu}
 \ee
where $R_\nu\equiv \rho_\nu/(\rho_\nu+\rho_\gamma)$ is the
fraction of the final radiation density in neutrinos, and
$R_\nu\approx0.41$ for $\xi_i\ll1$.

The amplitude of the CDM and baryon isocurvature perturbations,
calculated assuming $\zeta=\zeta_\gamma$, acquire an additional
term
\be
 \S_{{\rm cdm}/B}
 = 3(\zeta_{{\rm cdm}/B}-\zeta_\gamma) = 3(\zeta_{{\rm cdm}/B}-\zeta)
  + R_\nu\S_\nu 
 \,.
 \ee
 Hence for CDM/baryons created after curvaton decay (or in the
 inflaton scenario), for which $\tilde\zeta_{{\rm cdm}/B}=\zeta$, we
 now have a non-zero isocurvature matter perturbation
\be
 \S_{{\rm cdm}/B} = R_\nu \S_\nu \,,
 \ee
 for CDM/baryons created before curvaton decay, with
 $\tilde\zeta_{{\rm cdm}/B}=0$, we have
\be
 \S_{{\rm cdm}/B} = -3\zeta + R_\nu \S_\nu \,,
 \ee
 while for CDM/baryons created by curvaton decay, with
 $\tilde\zeta_{{\rm cdm}/B}=\zeta_\sigma$, we have 
%
\be 
\S_{{\rm cdm}/B} = 3 \left( {1-r\over r} \right) \zeta + R_\nu
   \S_\nu \,.  
\ee
If  $S_\nu\ll \zeta$, then
 $\zeta_\gamma$ can be identified to high accuracy with
$\zeta$,  and the changes in $\S_{{\rm cdm}/B}$ induced by the
 neutrino
isocurvature perturbations represent a small correction.

These changes do not correspond to any change in the evolution of the
CDM or baryons in the early Universe. They would be avoided if we
worked instead with the quantities $\hat \S_{\rm{cdm}/B} = 3(\tilde
\zeta_{\rm{cdm}/B} - \zeta)$, that define the perturbations on slices
of uniform total density (\eq{gud}). These quantities are independent
of $\S_\nu$ so that, for example, they always vanish if the CDM or
matter is created after curvaton decay with a density depending only
on the local energy density.  In a similar way, the quantities $\hat
\S_i= 3(\tilde \zeta_i - \zeta)$ reflect the early Universe situation
more directly than the quantities $\tilde S_i$, as we shall see when
discussing leptogenesis from curvaton decay. In all cases though, the
unhatted quantities are those commonly used at the `primordial' era,
as a starting point for the forward evolution of the perturbations.

\subsection{The residual neutrino isocurvature perturbation}

Now we wish to discuss the residual  neutrino isocurvature
perturbation in the curvaton scenario, along the same lines as for
the CDM and baryons.  To do this, we assume that the LMA solution
is correct so that the primordial lepton number densities have a
common value, determined by the total lepton number density that
is conserved in the early Universe. Since the lepton asymmetry is
small in that case, the neutrino isocurvature perturbation is
given to high accuracy by  \eq{finalSnu},
\be
 \S_\nu
  \approx
   {45\over7} \left({\xi\over\pi}\right)^2 \tilde\S_L \,,
 \ee
where the common asymmetry parameter $\xi$ satisfies the
nucleosynthesis constraint $|\xi|<0.07$.

To evaluate $\tilde\S_L$ we can follow closely the previous
discussions of CDM and baryon number. We take the epoch of
creation of lepton number $L$ to be the one at which this quantity
starts to be conserved. We assume that the lepton number
isocurvature perturbation at creation is zero or negligible,
except in the case that lepton number is created by the curvaton
decay. If lepton number is created  after curvaton decay,
$\tilde\S_L=0$ and there is no neutrino isocurvature perturbation.

If lepton number is created well before curvaton decay
and well before the curvaton dominates the density,
\be
 \tilde\S_L = -3\zeta_\gamma \,,
 \ee
giving
\be
 \S_\nu = - \frac{135}7 \( \frac{\xi}{\pi} \)^2 \zeta_\gamma \,.
\ee
For $|\xi|<0.07$ and using Eq.~(\ref{zetazetanu}), we have
\be
|S_\nu| < 0.01 \zeta
\,.
\ee
This is very small and may never by observable. 

Finally, if lepton number is created by out-of-equilibrium
curvaton decay,\footnote{The exact formula is $\hat\S_L
=3\zeta(1-r)/r$ where $\hat\S_L=3(\tilde \zeta_L -\zeta)$,
 but the difference is negligible in the regime
$(\xi/\pi)^2\ll 1$.}
\be
\tilde  \S_L \approx 3\(\frac{1-r}r\) \zeta \,.
\ee
If the curvaton does not dominate before it decays, this gives
\bea
 \S_\nu &\approx& 
\frac{135}7 \( \frac{1-r}r \) \( \frac\xi\pi \)^2 \zeta \\
 |\S_\nu| &<& 0.01 \frac{1-r}r \zeta  \lsim 15 \zeta \,, 
\eea
where the final inequality comes from the current bounds on the
non-Gaussianity parameter, $f\sub{NL}$, given in Eq.~(\ref{fnlpred}).
This upper bound represents a large anti-correlated neutrino
isocurvature component which may already be incompatible with current
data, though this possibility is yet to be tested against
observations.  If MAP further constrains the non-Gaussian parameter,
$f\sub{NL}$, (rather than detecting non-Gaussianity) the bound will
become $|\S_\nu| <0.2 \zeta$.

We note that the neutrino isocurvature {\em velocity} mode considered
by Bucher et al~\cite{BMT} is hard to generate in a curvaton scenario,
since we would expect any velocity perturbation left on large scales
after the curvaton decay to be suppressed by factors of order
$(k/aH)^2$ with respect to the (almost scale-invariant) density
perturbation, i.e., generated only by spatial gradients in the
curvaton field.


\section{Conclusions}

In this paper we have reviewed the mechanism by which a curvaton field
can generate a large-scale density perturbation after inflation, and
gone on to investigate the nature of the primordial perturbation that
is produced.
In many cosmological models the curvaton scenario can
reproduce the purely adiabatic and Gaussian density perturbation
(curvature perturbation) that is familiar from the inflaton scenario.
In another regime though, it can give large non-Gaussianity, and/or a
large `residual' isocurvature perturbation which is completely
correlated with the curvature perturbation and can in principle be
present in any or all of the baryonic matter, the CDM and the three
neutrino species.

The non-Gaussianity of the curvature perturbation $\zeta$ arises if
the curvaton fails to dominate the energy density before it decays.
The perturbation is described by a $\chi^2$-distribution, whose
non-Gaussianity is parameterised by
\be
f\sub{NL} =\frac5{4r}
\,,
\ee
where $r$ is approximately given by the curvaton density just before
decay, as a fraction of the total. 
In 2003, results from MAP satellite~\cite{MAP} will either detect this
non-Gaussianity or show that the curvaton density is at least a few
percent of the total when it decays.

The residual isocurvature perturbation in CDM or baryonic matter
arises if the CDM or baryon number is created either before curvaton
decay, or by the curvaton decay itself.  If it is created
significantly before curvaton decay
and before the curvaton dominates the energy density, 
the isocurvature perturbation is given by
\be
S\sub{cdm} = -3\zeta
\,,
\ee
and similarly for baryons.  This large, correlated isocurvature
perturbation is ruled out by observation for CDM.  In other words, CDM
cannot be created significantly before curvaton decay and before the
curvaton dominates.  This is a strong constraint on the cosmology,
implying for example that the curvaton scenario is inconsistent with
the CDM creation just after the end of inflation.  For baryons the
perturbation amplitude is close to current observational limits and
will be ruled out or detected in the near future, providing in the
latter case a smoking gun for the curvaton scenario.

If the CDM or baryon number is created by the curvaton decay itself,
\be
S\sub{cdm} = 3 \( \frac{1-r}r \) \zeta
\,,
\ee
and similarly for baryons. Unless $r$ is close to 1, this large
anti-correlated contribution is ruled out by observation for both CDM
and baryons. In other words, the curvaton decay cannot create CDM or
baryons unless the curvaton dominates the energy density before it
decays.

For the neutrino isocurvature perturbation we have presented
an analysis which, for the first time, includes the crucial issue
of lepton number, and is relevant for any cosmology. 
Although general formulae are given, we focus on the case where
the ratio of the neutrino and photon temperature is unperturbed,
because the late decoupling of neutrinos makes it very hard to see how
it could be otherwise. We also focus on the case that the e, $\mu$ and
$\tau$ lepton number densities are equal, which will be ensured by
mixing in the early Universe  at least if the large mixing angle solution
to the solar neutrino problem is correct.

With these assumptions, the isocurvature {\em energy} density
perturbation $S_\nu$ of massless neutrinos is related to the lepton {\em
  number} density isocurvature perturbation $\tilde\S_L$ by
\be
\S_\nu = \frac{45}7 \(\frac{\xi}{\pi} \)^2 \tilde\S_L \\
\,, 
\ee 
where $\xi$ is the lepton asymmetry parameter, related to the lepton
number density by $\xi= 4.02 (n_L/n_\gamma)$.  If the lepton number
density is of order the baryon number density as is usually supposed,
$|\xi| \sim 10^{-9}$ and  the neutrino isocurvature perturbation will
be completely undetectable.  Observationally though, the lepton
density is subject only to the nucleosynthesis bound $|\xi|<0.07$,
which may allow a detectable perturbation. However neutrino
isocurvature perturbations in the presence of a significant lepton
asymmetry have never been tested against observations.

In the curvaton scenario, the formulas for the residual isocurvature
perturbation $\tilde\S_L$ are the same as those for the CDM and baryon
perturbations.  If leptogenesis occurs after curvaton decay there is
no residual neutrino isocurvature perturbation.  If it is
significantly before curvaton decay
and before the curvaton dominates, 
$\tilde\S_L = -3\zeta$ and $|\S_\nu| < 0.01\zeta$ which may never be
observable. The interesting result comes in the third case, that the
curvaton decay itself causes leptogenesis.  Then $\tilde\S_L=
3((1-r)/r)\zeta$, and if the curvaton does not dominate before decay
we have a neutrino isocurvature perturbation whose magnitude is
related to the non-Gaussianity parameter $f\sub{NL}=5/4r$,
\be
\S_\nu \approx \frac{135}7 \(\frac{\xi}{\pi} \)^2 \(
{4\over5}f\sub{NL}-1 \)  \zeta 
\,.
\ee
If the present bound on non-Gaussianity, $|f\sub{NL}| < 2000$, is
saturated this permits a huge effect, presumably already ruled out by
observation, and even the expected MAP bound on $|f\sub{NL}|<20$ will
still allow a significant effect. Observational bounds on
$f\sub{NL}$ and $\S_\nu$ in this scenario should be obtained jointly,
taking into account the correlation of these two quantities with each
other and with the curvature perturbation $\zeta$.

In summary, we have shown that different curvaton scenarios offer
a number of distinctive observational predictions which may be tested
by forthcoming experiments.

{\em Note added:}
A related paper by Moroi and Takahashi~\cite{MT2} discussing the case
of matter isocurvature perturbations in the curvaton scenario appeared
while this work was in progress.

\acknowledgments We are grateful to Kostas Dimopoulos, Luca Amendola
and Martin Bucher for useful discussions. This work was supported in
part by PPARC grants PPA/G/S/1999/00138, PPA/G/S/2000/00115 and
PGA/G/O/2000/00466, and by EU grant HPRN-CT-2000-00152.  DW is
supported by the Royal Society.


\end{document}